\documentclass[useAMS,usenatbib]{mn2e}

\usepackage{graphicx}

\title[Flickering of KR\,Aur and UU\,Aqr]{Flickering study of nova like systems KR\,Aur and UU\,Aqr}
\author[A. Dobrotka, S. Mineshige and J. Casares]{A. Dobrotka$^{1,2}$\thanks{E-mail: andrej.dobrotka@stuba.sk}, S. Mineshige$^{2}$\thanks{E-mail: shm@kusastro.kyoto-u.ac.jp}, J. Casares$^{3,4}$\thanks{E-mail: jcv@iac.es}\\
$^1$Department of Physics, Institute of Materials Science, Faculty of Materials Science and Technology, Slovak University\\ of Technology in Bratislava, J\'ana Bottu 25, 91724 Trnava, Slovak Republic\\
$^2$Department of Astronomy, Graduate School of Science, Kyoto University, Sakyo-ku, Kyoto 606-8502, Japan\\
$^3$ Instituto de Astrof\'{\i}sica de Canarias, E-38200 La Laguna, Tenerife, Spain\\
$^4$ Departamento de Astrof\'{\i}sica, Universidad de La Laguna, Avda. Astrof\'{\i}sico Francisco S\'anchez s/n, E-38271 La Laguna, Tenerife, Spain
}

\begin{document}

\date{Accepted ???. Received ???; in original form \today}

\pagerange{\pageref{firstpage}--\pageref{lastpage}} \pubyear{2011}

\maketitle

\label{firstpage}

\begin{abstract}
We present a study of the flickering activity in two nova like systems KR\,Aur and UU\,Aqr. We applied a statistical model of flickering simulations in accretion discs based on turbulent angular momentum transport between two adjacent rings with an exponential distribution of the turbulence dimension scale. The model is based on a steady state disc model which is satisfied in the case of hot ionized discs of nova like cataclysmic variables. Our model successfully fits the observed power density spectrum of KR\,Aur with the disc parameter $\alpha = 0.10 - 0.40$ and an inner disc truncation radius in the range $R_{\rm in} = 0.88 - 1.67 \times 10^9$\,cm. The exact values depend on the mass transfer rate in the sense that $\alpha$ decreases and $R_{\rm in}$ increases with mass transfer rate. In any case, the inner disc radius found for KR\,Aur is considerably smaller than in quiescent dwarf novae, as predicted by the disc instability model. On the other hand, our simulations fail to reproduce the power density spectrum of UU\,Aqr. A tantalizing explanation involves the possible presence of spiral waves, which are expected in UU\,Aqr, because of its low mass ratio, but not in KR\,Aur. In general our model predicts the observed concentration of flickering in the central disc. We explain this by the radial dependence of the angular momentum gradient.
\end{abstract}

\begin{keywords}
stars: cataclysmic variables - stars: individual: KR\,Aur, UU\,Aqr - accretion: accretion discs
\end{keywords}

\section{Introduction}
\label{introduction}

Cataclysmic variables are interacting binaries with a late type secondary star overfilling its Roche lobe and transferring mass from the inner Lagrangian point $L_1$ toward a white dwarf primary. When the white dwarf in not magnetic (or weakly magnetic i.e. an intermediate polar) an accretion disc is formed. Disc matter is transported inwards whereas angular momentum is transported outwards. Accretion is a common phenomenon of a broad family of astrophysical objects. X-ray binaries are the most similar accreting systems to cataclysmic variables. They possess a black hole or a neutron star as central objects. Supermassive black holes are also the central engine which powers active galactic nuclei. Accretion of matter through a disc is the source of brightness variations on a wide range of time scales and energies. The time scale of the variability scales with the mass of the compact object and this is the reason why cataclysmic variables provide excellent laboratories to study accretion physics.

The most characteristic signature of accretion is the flickering activity. These brightness variations, with amplitudes in the range of $0.1 - 1$ magnitude, last from seconds to tens of minutes. There are four basic physical interpretations for the flickering activity that we know of; 1) interaction of the gas stream from the secondary with the outer edge of the disc , 2) turbulent flow in the accretion disc, 3) magnetic dissipation/reconnection in the disc corona and 4) interaction of the inner edge of the disc with the central white dwarf, the so-called boundary layer. Furthermore, a cellular-automaton model was proposed by \citet{yonehara1997} to describe the timing properties of flickering in cataclysmic variables. In this model light fluctuations are produced by occasional flare like events and subsequent avalanche flow in the accretion disc atmospheres. Flares are assumed to be triggered when the mass density exceeds a critical limit. There have been several attempts to localize the source of flickering. For example, enhanced flickering activity during the orbital hump before the eclipse of U\,Gem was observed by \citet{warner1971}. This allowed them to identify the flickering source with the interaction of the gas stream with the outer edge of the disc (hot spot). On the other hand, \citet{bruch1992} studied energies and variability time scales in a large sample of cataclysmic variable and symbiotic systems and concluded that both the disc and the boundary layer are responsible for the flickering. The combination of the inner disc and hot spot is the source of flickering variability in the case of Z\,Cha \citep{bruch1996} and HT\,Cas, V2051\,Oph, UX\,UMa, IP\,Peg \citep{bruch2000}. Eclipse mapping of V2051\,Oph identified the inner disc as the source of high frequency flickering and the stream overflow as the source of low frequency variability \citep{baptista2004}. The inner disc is also responsible for the flickering in the symbiotic system T\,CrB \citep{zamanov1998}. This was studied and confirmed by statistical modeling of turbulent accretion disc flow by \citet{dobrotka2010}. The authors explained the central concentration of flickering by the large number of turbulent eddies required to transfer angular momentum from the inner disc radius outwards. Therefore, the accretion disc appears as the most common source of flickering in cataclysmic variables and related object such as symbiotic systems, specially its inner parts and boundary layer.

Nova likes are a subclass of cataclysmic variables (see \citealt{warner1995} for review) with high mass transfer rates. Their discs are hot and fully ionized most of the time which means that the mass accretion rate is constant through the whole disc (i.e. steady state). This is opposite to dwarf novae where the mass transfer rate is lower, the mass accretion rate is not constant and intersects the critical value of thermal stability somewhere in the disc at some time. This drives a viscous-thermal instability which triggers the dwarf novae phenomenon (see \citealt{kato1998}, \citealt{lasota2001} for review). Dwarf novae discs are not in steady-state and switch between hot (ionized) and cold (non ionized) states with the former presenting smaller disc truncation radii. A small inner disc truncation radius is expected also in the case of nova like systems. The inner disc truncation radius can be produced either by weak magnetic field (i.e. not strong enough to qualify the system as an intermediate polar) or some kind of evaporation (see e.g. \citealt{meyer1994}). The case of ionized steady state discs is well described by the \citet{shakura1973} accretion disc model.

Extensive work has been devoted to localize the source of flickering in cataclysmic variables (see e.g. \citealt{bruch1996}, \citealt{zamanov1998}, \citealt{bruch2000}, \citealt{baptista2004}), but we know little about the physical phenomenon responsible for its variability. The location of flickering in the inner disc and boundary layer incorporates three possible physical processes: turbulence, magnetic reconnection and interaction of the inner disc with the central star. In this paper we are trying to test whether turbulence is the main mechanism responsible for flickering by applying the statistical model of turbulent accretion flow developed in \citet{dobrotka2010}. The model is applied to the flickering activity observed in the nova likes KR\,Aur and UU\,Aqr. In Sec.~\ref{systems} we summarize the observational flickering properties of KR\,Aur and UU\,Aqr. In Sec.~\ref{modelling} we retrace the statistical turbulence model of Dobrotka et al. (2010). The uncertainty of the measured parameters is discussed in Sec.~\ref{errors}. The final results are presented in Sec.~\ref{results}, discussed in Sec.~\ref{discussion} and summarized in Sec.~\ref{summary}.

\section{Observational results}
\label{systems}

\subsection{KR\,Aur}
\label{kraur}

KR\,Aur is a nova like system of VY\,Scl type. Its flickering activity was studied by \citet{kato2002} during a high state. The authors used 17 observations, with a mean coverage of 9000\,s and 50\,s time resolution, to calculate the  power density spectra (PDS). The PDS of the flickering is well described by a broken power law, characteristic of red noise. The power law has a slope with index of -1.63 and a cut off frequency at 31.6\,d$^{-1}$ (or 1.5\,d$^{-1}$ in logarithmic units).

Quasi-periodic variability was detected on some nights with a characteristic time scale of $10 - 15$\,min. No superhumps or other coherent signals were observed.

\subsection{UU\,Aqr}
\label{uuaqr}

Photometric light curves of the nova like UU\,Aqr were obtained by \citet{baptista2008} with 10\,s time resolution and a mean duration of 8000\,s per run. The PDS shows a power law with index -1.5 and a break frequency at $\sim 12.96$\,d$^{-1}$ (or 1.11 in logarithmic units). The eclipse mapping technique was applied to analyze the short term variability. Low and high frequency flickering maps are dominated by emission from two asymmetric arcs similar to the case of the dwarf nova IP\,Peg \citep{baptista2002}. The interpretation of the flickering activity in UU\,Aqr is then similar to IP\,Peg, i.e. tidally-induced spiral shock waves in the outer regions of a large accretion disc. Baptista et al. (2002) have suggested that the flickering in UU\,Aqr is caused by turbulence generated by the disc-spiral wave interaction.

\section{Flickering modeling and simulations}
\label{modelling}

\citet{dobrotka2010} developed a statistical method to model the flickering light curves using turbulence. The method is based on angular momentum transport between two adjacent concentric rings in a disc. The angular momentum is transported by discrete turbulent spherical bodies with a dimension scale $x$ described by an exponential distribution function. The dimension scale is proportional to the radial propagation of the turbulent body in the accretion disc. The maximum dimension is the scale height of the disc, while the minimum is limited by zero.

The basic idea is that the sum of the angular momentum difference $\Delta l(r)$ (where $r$ is the distance from the disc center) of all spherical bodies in two adjacent rings must be equal to the global angular momentum difference between the two rings $\Delta L(r)$, i.e. $\Delta l(r) \sim \Delta L(r)$. Since the scale height, density, radial and tangential velocity of the matter in the rings is a function of $r$, the angular momentum differences must be correlated with a certain parameter $k(r)$ which depends on the radial position of every pair of rings, and is given by
\begin{equation}
k(r) = \frac{\Delta L(r)}{\Delta l(r)}.
\label{correlation_coef}
\end{equation}
Therefore, $k(r)$ is a measure of the quantity of events between two adjacent rings at radial distance $r$. This parameter is needed to estimate the number of flares with duration $t \sim x/v_{\rm r}(r)$, where $v_{\rm r}(r)$ is the radial viscous velocity. The number of events with a certain duration is summed up through all pair of rings at distances $r_{\rm i}$
\begin{equation}
\zeta(t) = \sum\limits_{i=1}^{n} k(r_{\rm i}) f(r_{\rm i},t).
\end{equation}
where $f(r_{\rm i},t)$ is the exponential distribution of turbulent eddy sizes with $t \sim x/v_{\rm r}(r)$ at distance $r_{\rm i}$ from the disc center and $n$ is the number of ring pairs. By selecting the number of flares with durations $t$ between a minimal and maximal value and a sampling time step, we get the final histogram used in the synthetic light curve generation (see \citealt{dobrotka2010} for details).

The disc radial profile is based on the \citet{shakura1973} model. Therefore, the disc must be in the hot ionized steady state, characteristic of dwarf novae outbursts (see \citealt{lasota2001} for review). Nova like systems are adequate for such modeling because of their high mass transfer rates which tend to keep them in the hot branch of dwarf novae activity (see \citealt{warner1995} for review).

We used this method to test whether the turbulence scenario is able to fit the observed PDS with realistic values of the disc $\alpha$ parameter and the inner disc truncation radius. The outer disc radius $R_{\rm out}$ is taken as half\footnote{The tidal perturbation of the secondary truncates the outer disc radius at $\sim 0.9 \times R_{\rm L1}$ and hence provides an upper limit to $R_{\rm out}$, but our simulations are not sensitive to this parameter (see \citealt{dobrotka2010})} radius of the primary Roche lobe $R_{\rm L1}$ \citep{paczynski1971};
\begin{equation}
R_{\rm L1} = 0.462~a~\left( \frac{M_{\rm 1}}{M_{\rm 1} + M_{\rm 2}} \right)^{1/3}\,({\rm cm}),
\end{equation}
where $M_{\rm 1}$ and $M_{\rm 2}$ are the primary and secondary mass respectively and $a$ is the distance between the two stars calculated from Kepler's Third law
\begin{equation}
a = 3.53 \times 10^{10} \left( \frac{M_{\rm 1}}{{\rm M_{\rm \odot}}} \right)^{1/3} (1 + q)^{1/3} \left( \frac{P_{\rm orb}}{1\,{\rm h}} \right)^{2/3}\,({\rm cm}),
\end{equation}
$q = M_{\rm 2} / M_{\rm 1}$ is the mass ratio and $P_{\rm orb}$ the orbital period.

The critical mass transfer rate for a disc to be in a hot stage is given by equation (\citealt{hameury1998})
\begin{equation}
\dot{M}_{\rm tr} = 9.5 \times 10^{15} \alpha^{0.01} \left( \frac{M_1}{{\rm M_{\odot}}} \right)^{-0.89} \left( \frac{r}{10^{10}\,{\rm cm}} \right)^{2.68}\,{\rm g}\,{\rm s^{-1}}.
\end{equation}
Using $\alpha$ values from 0.01 to 1.00 we obtain a critical mass transfer rates of $7.8-8.2  \times 10^{16}$\,g\,s$^{-1}$ for both KR\,Aur and UU\,Aqr. By increasing the outer disc radius to $\sim0.9 \times R_{\rm L1}$ the critical mass transfer rate rises to $3.8 - 4.0 \times 10^{17}$\,g\,s$^{-1}$ for KR\,Aur and $3.8 - 3.9 \times 10^{17}$\,g\,s$^{-1}$ for UU\,Aqr. Therefore, we decided to take $8 \times 10^{16}$\,g\,s$^{-1}$ as a lower limit to the mass transfer rate for a steady disc. Simulations were also computed for $1 \times 10^{17}$, $5 \times 10^{17}$ and $1 \times 10^{18}$\,g\,s$^{-1}$. The model input parameters and the PDS values, as reported in the literature, are summarized in Table~\ref{system_parameters}.
\begin{table}
\caption{Measured and calculated parameters. $M_1$ is the primary mass in solar units, $q$ is the mass ratio, $pl$ index is the power law index as measured from the PDS, $f_{\rm ctf}$ the cut off frequency in mHz, $P_{\rm orb}$ the orbital period in hours, $R_{\rm wd}$ the white dwarf radius estimated from \citet{nauenberg1972} in $10^9$\,cm and $R_{\rm out}$ the outer disc radius as half of the primary Roche lobe in units of $10^{10}$\,cm.}
\begin{center}
\begin{tabular}{lcccr}
\hline
\hline
Object & $M_1$ & $q$ & $pl$ index & log$(f_{\rm ctf})$ \\
 & (M$_{\rm \odot}$) & & & (d$^{-1}$) \\
\hline
KR\,Aur & 0.59$^{1}$ & 0.6$^{1}$ & -1.63$^{2}$ & 1.50$^{2}$ \\
UU\,Aqr & 0.67$^{3}$ & 0.3$^{3}$ & -1.50$^{4}$ & 1.11$^{4}$ \\
\hline
\hline
Object & $P_{\rm orb}$ & $R_{\rm wd}$ & $R_{\rm out}$ & \\
 & (h) & ($10^9$\,cm) & ($10^{10}$\,cm) & \\
\hline
KR\,Aur & 3.907$^{1}$ & 0.87 & 1.87 & \\
UU\,Aqr & 3.900$^{3}$ & 0.80 & 1.95 & \\
\hline
\end{tabular}
\end{center}
References:\\
1 - \citet{shafter1983}, 2 - \citet{kato2002}, 3 - \citet{baptista1994}, 4 - \citet{baptista2008}
\label{system_parameters}
\end{table}

As mentioned above the dimensionless parameter $\alpha$ and the inner disc radius $R_{\rm in}$ are the free parameters of our simulation. We calculated PDS power law indexes ($pl$ hereafter) and cut off frequencies ($f_{\rm ctf}$ hereafter) for the parametric space $\alpha - R_{\rm in}$ with a resolution of $20 \times 20$ using the Lomb-Scargle method \citep{scargle1982}. The $\alpha$ parameter is allowed to vary from 0.01 up to 1.00 which are typical values for cataclysmic variables (see \citealt{warner1995}, \citealt{lasota2001} for review), whereas $R_{\rm in}$ varies from the vicinity of the white dwarf radius (i.e. 1 percent larger) up to $2.0 \times 10^9$\,cm in 20 steps. The white dwarf radius is estimated from the formula given in \citet{nauenberg1972}. For every combination of parameters we calculated a synthetic light curve and its PDS. The synthetic light curves were sampled in identical manner as the observations of KR Aur and UU Aqr reported in \citet{kato2002} and \citet{baptista2008}. These define the frequency limits 
and the resolution of the synthetic PDS. We repeated the simulation 1000 times and then calculated the mean PDS. The resulting mean PDS is binned into 20 equally spaced bins and fitted with two power laws. The break frequency $f_{\rm ctf}$ and the red noise slope $pl$ are the main output parameters.

\section{Uncertainty estimate}
\label{errors}

Unfortunately we have no information about the measurement errors of the observed PDS parameters used in this work. The uncertainty must be taken into account because of the limited number of data used in the mean PDS calculation. \citet{kato2002} only used 17 observations while \citet{baptista2008} used 31. This is considerably lower than the 1000 synthetic runs used in our simulations. Furthermore, the length of the observed light curves is variable and some are very short. Our simulated data sets have a constant duration in order to get confident PDS parameters. This can have an impact in the observed PDS properties and must be taken into account in the analysis of the deviation between the observed and simulated data.

\begin{figure*}
\includegraphics[width=60mm,angle=-90]{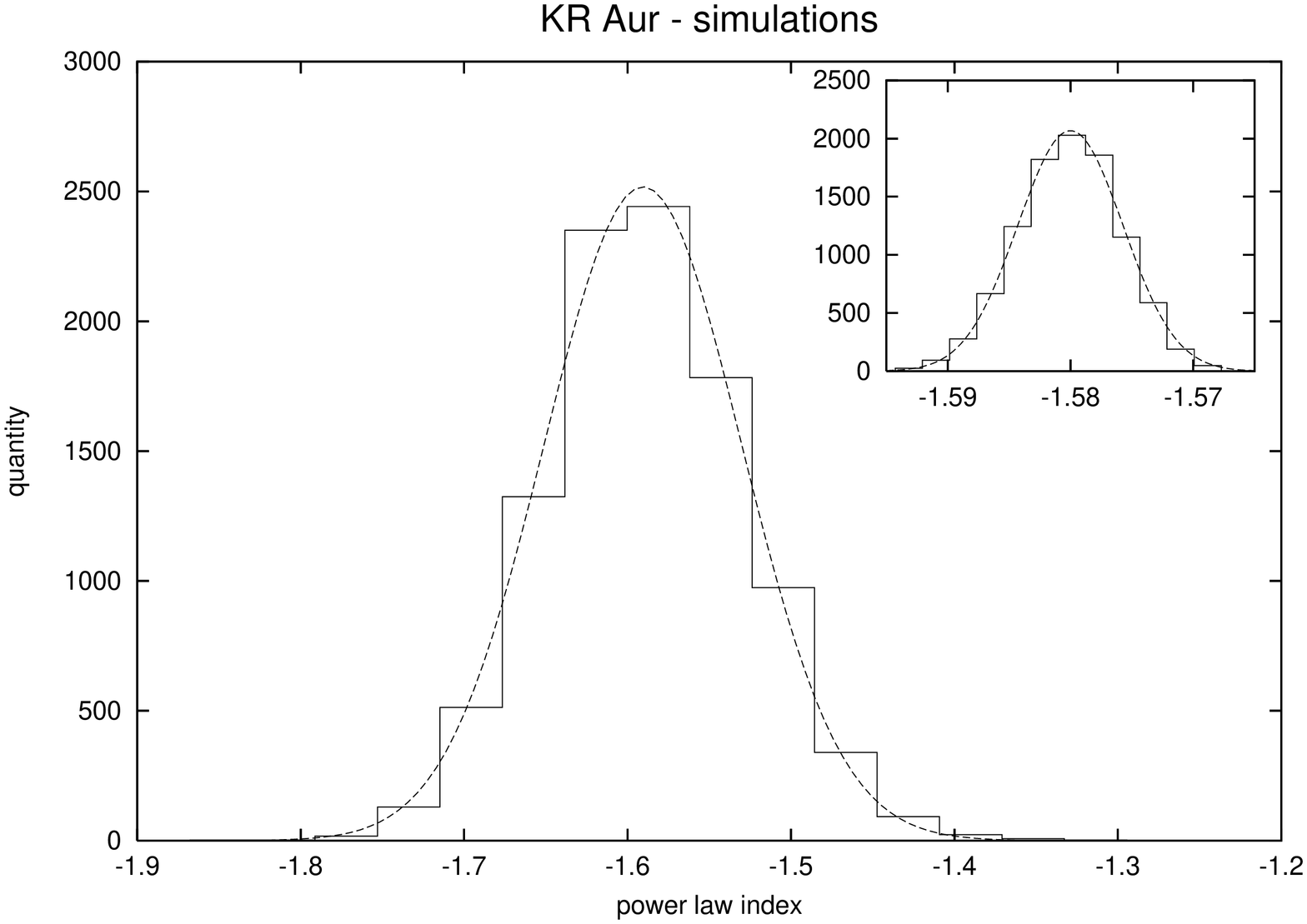}
\includegraphics[width=60mm,angle=-90]{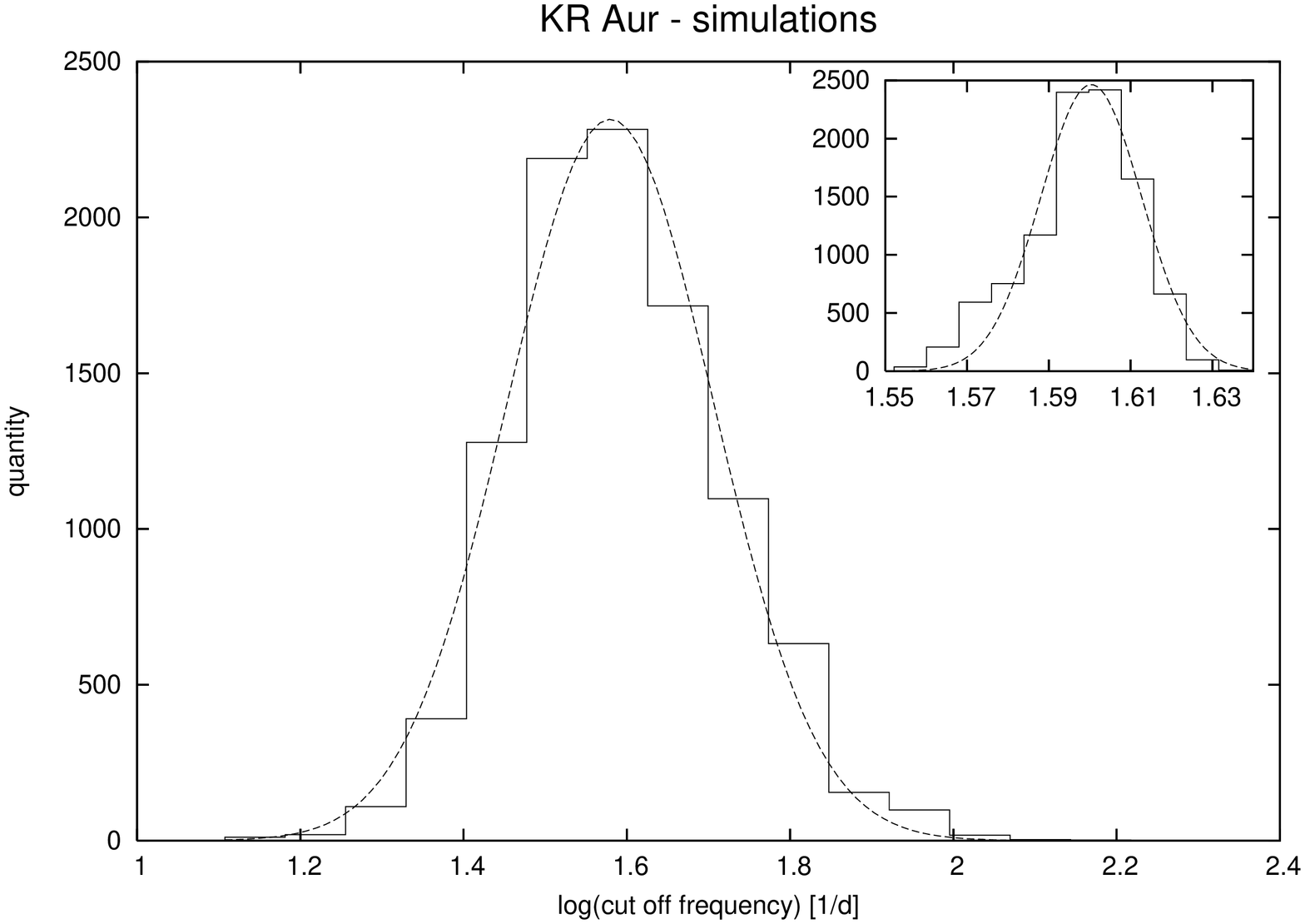}\\
\includegraphics[width=60mm,angle=-90]{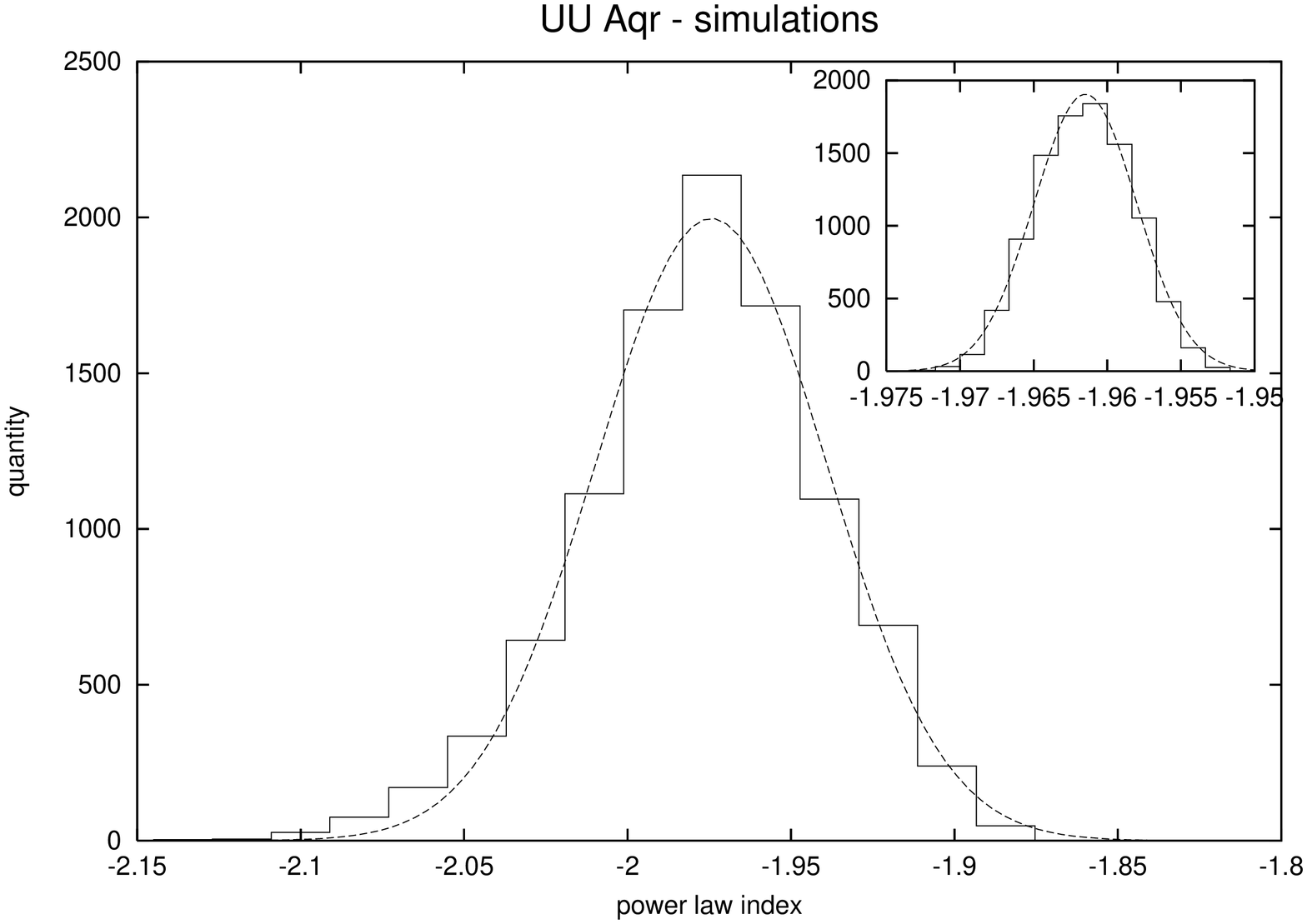}
\includegraphics[width=60mm,angle=-90]{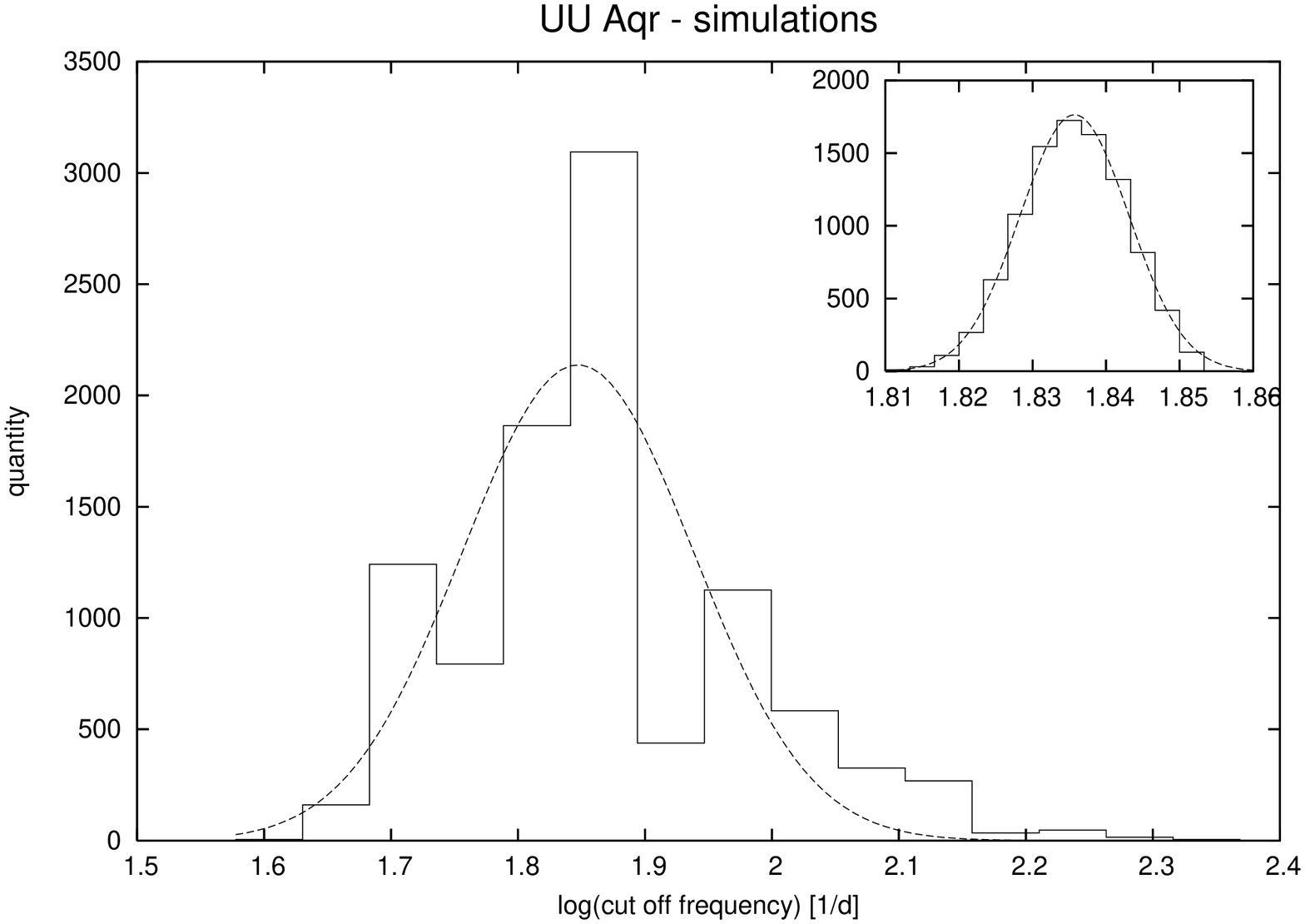}
\caption{Histograms of the simulated PDS parameters obtained using an ad hoc model with $\dot{M}_{\rm tr} = 1.0 \times 10^{17}$\,g\,s$^{-1}$, $\alpha$ = 0.20 and $R_{\rm in} = 1.0 \times 10^9$\,cm. Left panels show the histograms of the power law index for KR\,Aur (top) and UU\,Aqr (bottom) whereas right panels present the histograms of the cut off frequency in logarithm  units (also KR\,Aur on top and UU\,Aqr at the bottom). The dotted lines show the Gaussian fits to the histograms. Seventeen synthetic runs were produced to generate the PDS of KR\,Aur and 31 runs for UU\,Aqr. The histograms in the inset panels were produced using 1000 synthetic runs. The simulations were repeated 10000 times in each case.}
\label{simul_hist}
\end{figure*}

In order to estimate this, we performed simulations of 17 synthetic light curves for KR\,Aur with identical duration and sampling as described in Sec.~\ref{kraur} using an ad hoc model with $\dot{M}_{\rm tr} = 1.0 \times 10^{17}$\,g\,s$^{-1}$, $\alpha$ = 0.20 and $R_{\rm in} = 1.0 \times 10^9$\,cm. A mean PDS was calculated from the synthetic light curves and the process was repeated 10000 times. The resulting histograms of the power law index and the logarithm of the cut off frequency are depicted on the upper two panels of Fig.~\ref{simul_hist}, together with their best Gaussian fits. The simulated PDS parameters cluster at the mean values $pl = -1.59$ and ${\rm log}(f_{\rm ctf}) = 1.58$\,d$^{-1}$ with 1-$\sigma$ of 0.06 and 0.12\,d$^{-1}$ respectively. We take this 1-$\sigma$ values as our estimate of the error in the measured PDS parameters. For comparison, we obtain $pl = -1.580 \pm 0.004$ and ${\rm log}(f_{\rm ctf}) = 1.60 \pm 0.01$\,d$^{-1}$ when increasing the number of synthetic light curves to 1000. The corresponding histograms are shown in the inset panels to Fig.~\ref{simul_hist}. A similar analysis was also performed for the case of UU\,Aqr using 31 synthetic light curves per mean PDS as in the observations of \citet{baptista2008}. The duration and resolution of the synthetic light curves is identical to the mean values quoted in Sec.~\ref{uuaqr}. The PDS parameters cluster at the mean values $pl = -1.97$ and ${\rm log}(f_{\rm ctf}) = 1.85$\,d$^{-1}$ with 1-$\sigma$ widths of 0.04 and 0.09\,d$^{-1}$ respectively. For comparison 1000 synthetic light curves per mean PDS yields $pl = -1.961$ and ${\rm log}(f_{\rm ctf}) = 1.836$\,d$^{-1}$ with 1-$\sigma$ widths of 0.004 and 0.007\,d$^{-1}$ respectively. The histograms of the PDS parameters for UU\,Aqr and their best Gaussian fits are depicted in the lower panels of Fig.~\ref{simul_hist}.

\section{Results}
\label{results}

Fig.~\ref{results_grid} presents the simulated PDS parameters in the parametric space $\alpha - R_{\rm in}$ using a grey scale. Only the case for $\dot{M}_{\rm tr} = 8 \times 10^{16}$\,g\,s$^{-1}$ is shown for illustration. The isocontours indicate the measured values (middle line) and the $\pm 1$-$\sigma$ errors as estimated in Sec.~\ref{errors}. No contours are visible in the case of UU\,Aqr because they lie outside the plotted $\alpha$ parameter range $0.01 - 1.00$. In this paper we are mainly interested in the $\alpha$ range $0.1 - 0.5$. The lower limit is set by the minimum $\alpha$ parameter required by the disc instability model (DIM) for hot discs while the upper limit is somehow arbitrary. Values higher than 0.5 are theoretically possible but previous works suggest that they are unlikely (see e.g. \citealt{schreiber2003}, \citealt{schreiber2004}, \citealt{king2007}). Simulations of hot discs in cataclysmic variables typically yield $\alpha = 0.1 - 0.2$ whereas observations constrain $\alpha$ to the range $0.1 - 0.4$.

\subsection{KR\,Aur}
\label{results_kraur}

\begin{figure*}
\includegraphics[width=52mm,angle=-90]{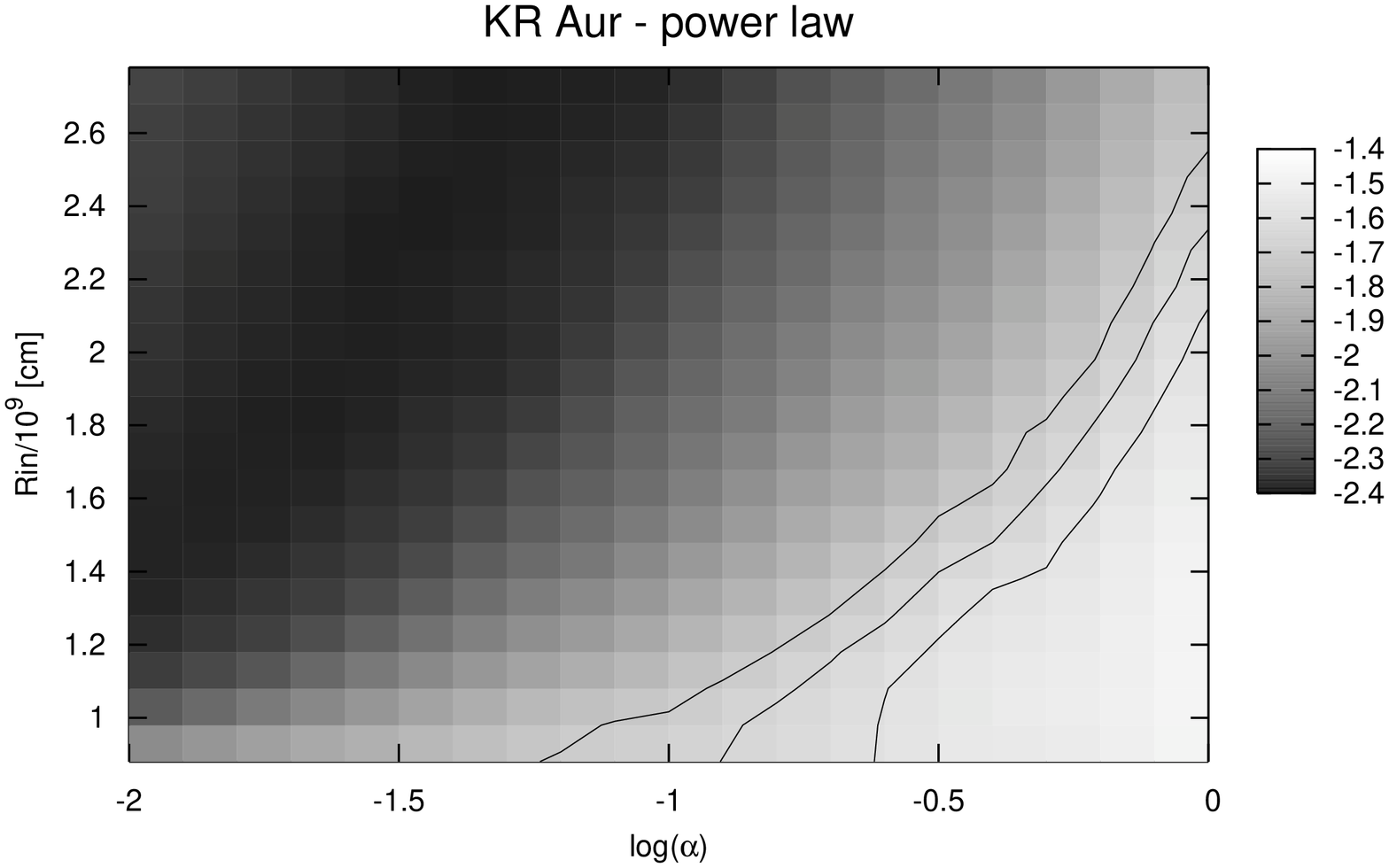}
\includegraphics[width=52mm,angle=-90]{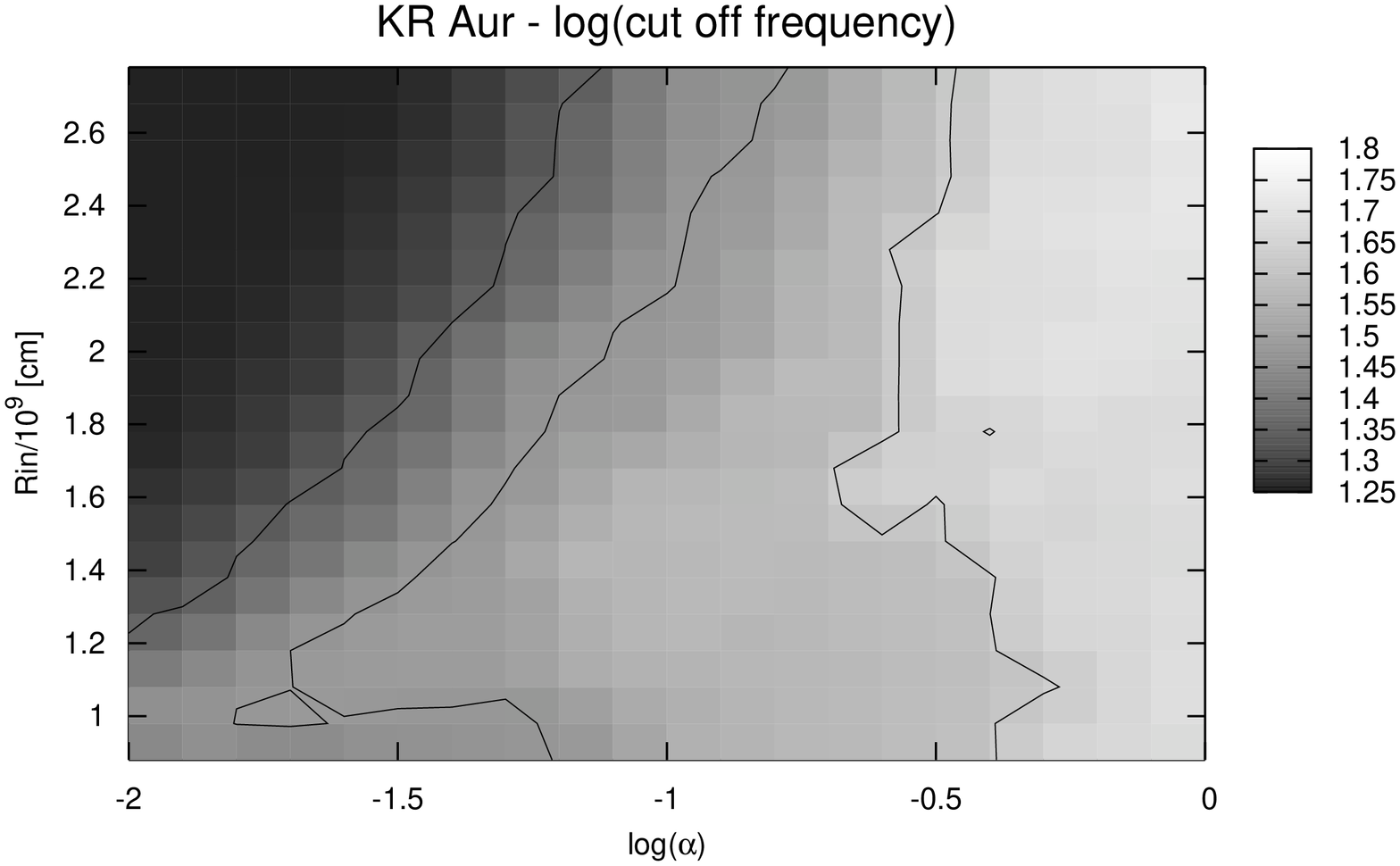}\\
\includegraphics[width=52mm,angle=-90]{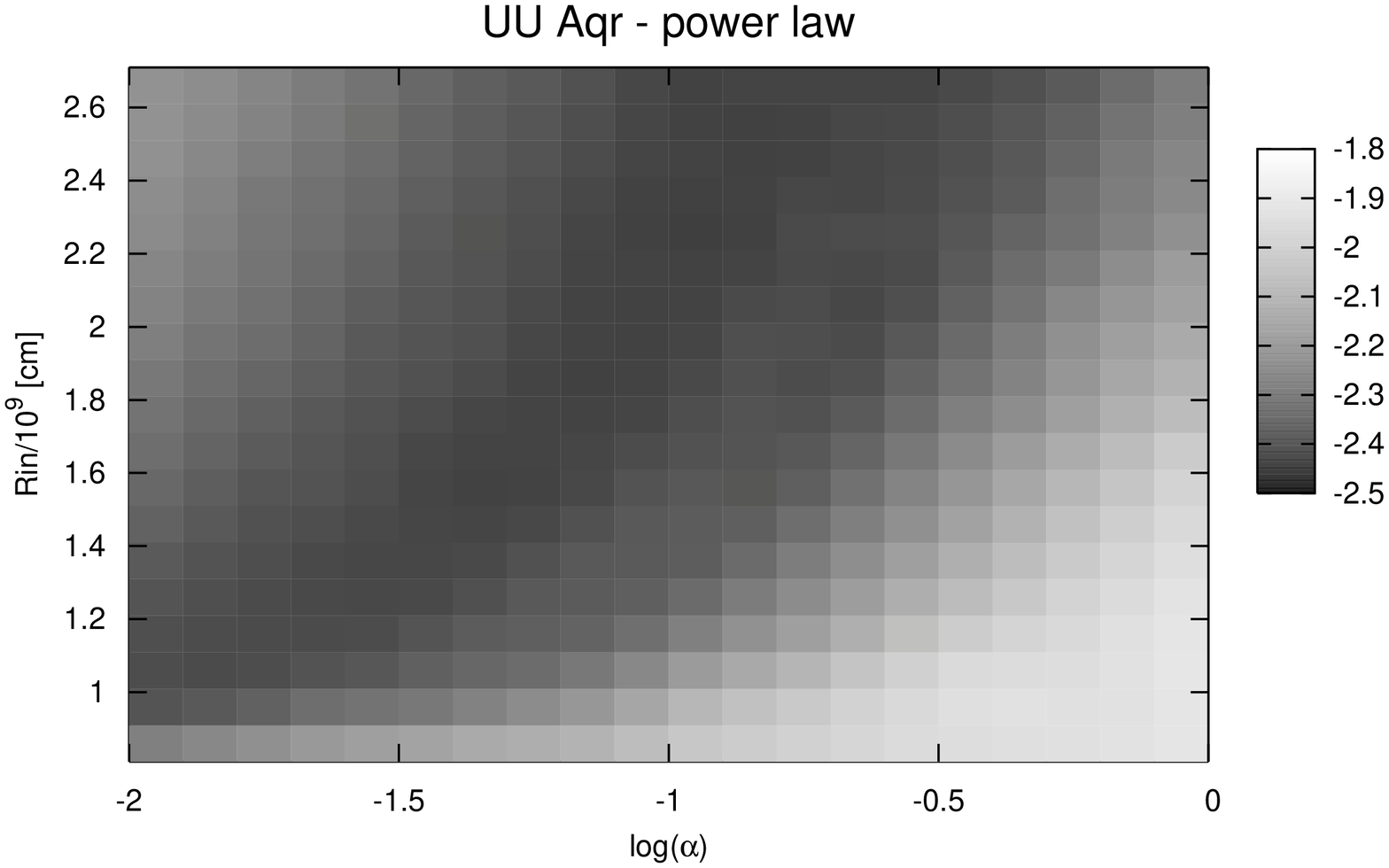}
\includegraphics[width=52mm,angle=-90]{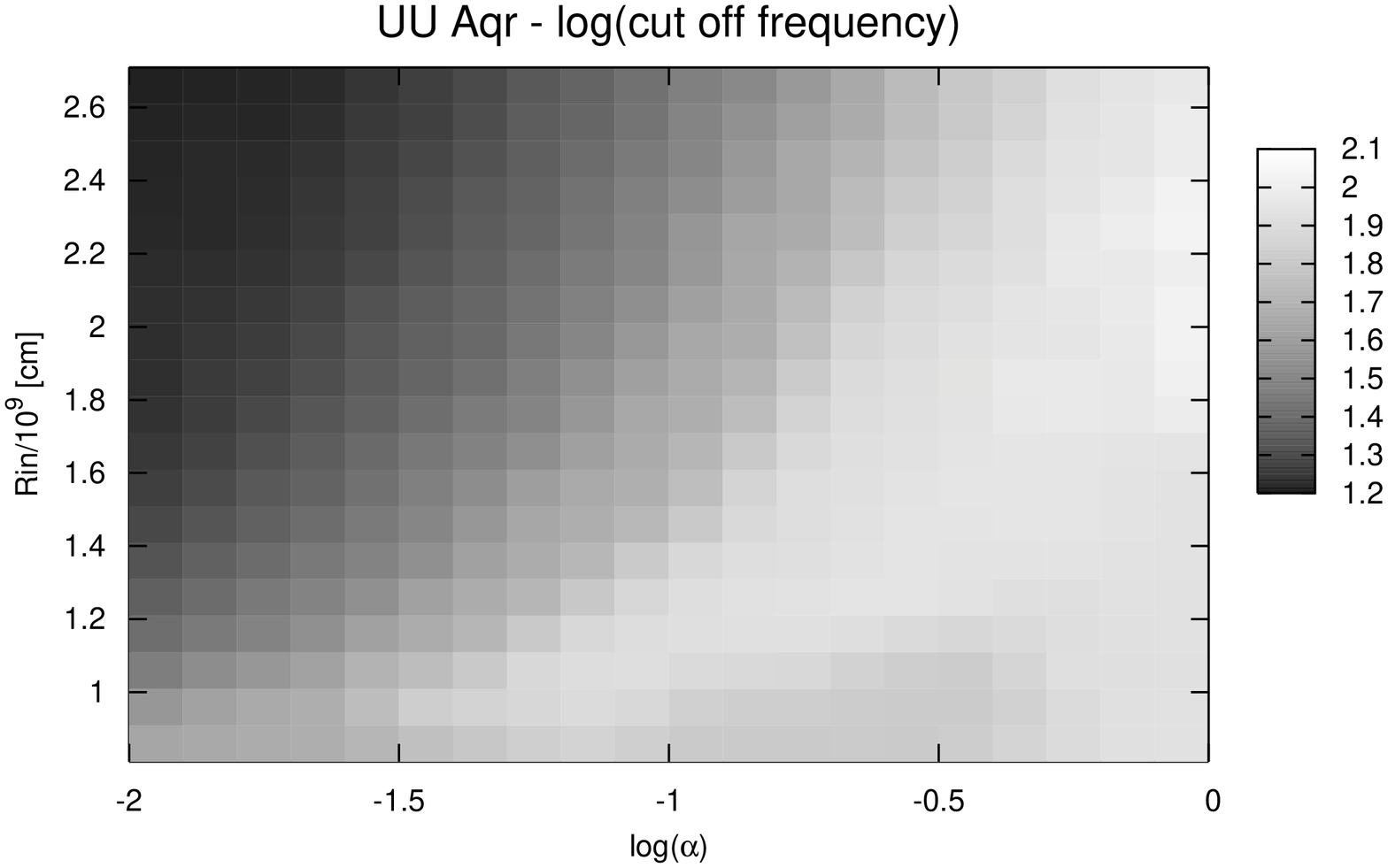}
\caption{PDS parameters for KR\,Aur (upper panels) and UU\,Aqr (lower panels) in the parametric space $\alpha - R_{\rm in}$ for $\dot{M}_{\rm tr} = 8 \times 10^{16}$\,g\,s$^{-1}$. Left panels represent the fitted power law and the right panels the logarithm of the cut-off frequency Contour lines indicate the mean value of the parameters and their $\pm 1$-$\sigma$ limits. For UU\,Aqr these lie outside the calculated range $\alpha = 0.01 - 1.00$.}
\label{results_grid}
\end{figure*}

\begin{figure*}
\includegraphics[width=60mm,angle=-90]{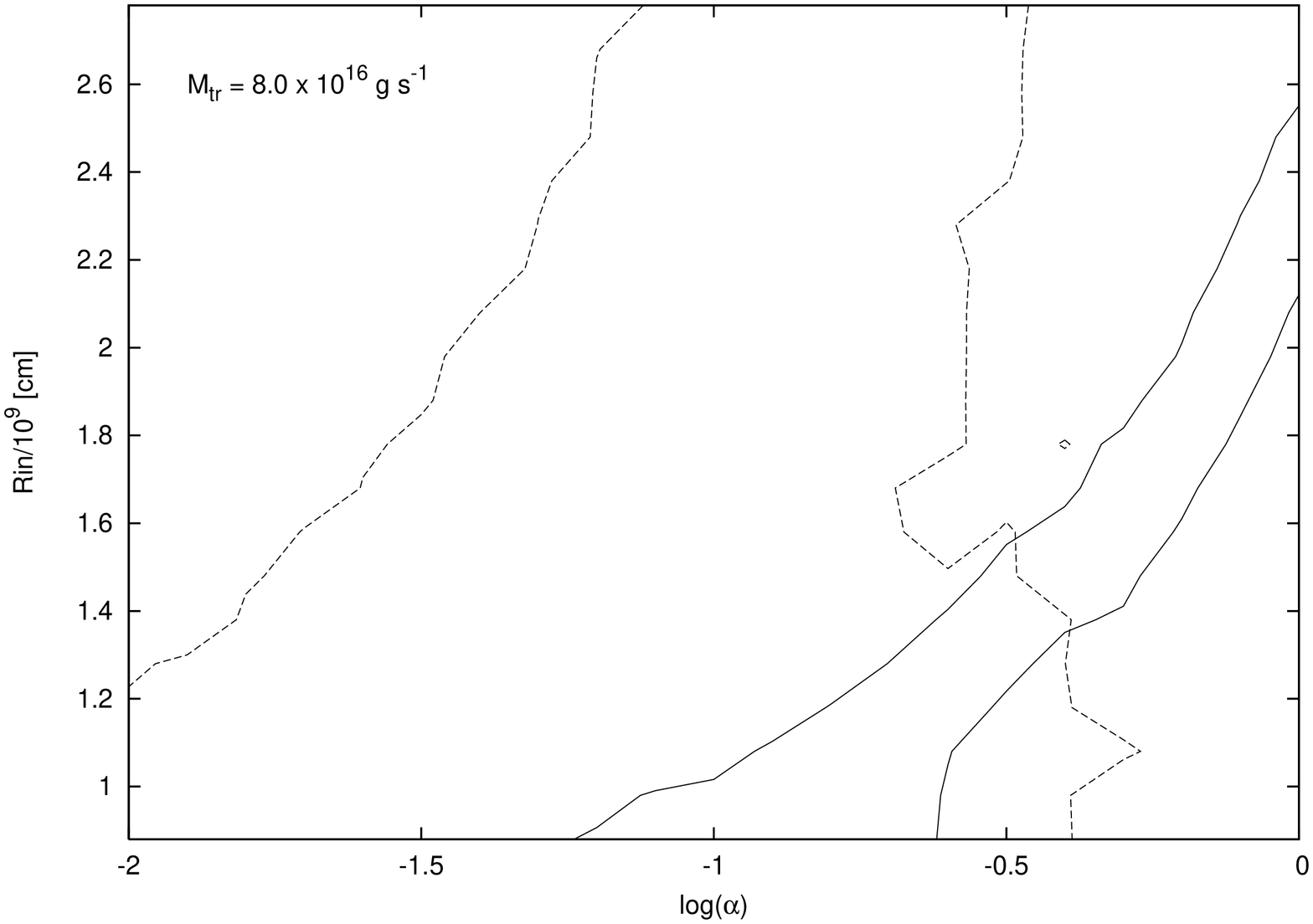}
\includegraphics[width=60mm,angle=-90]{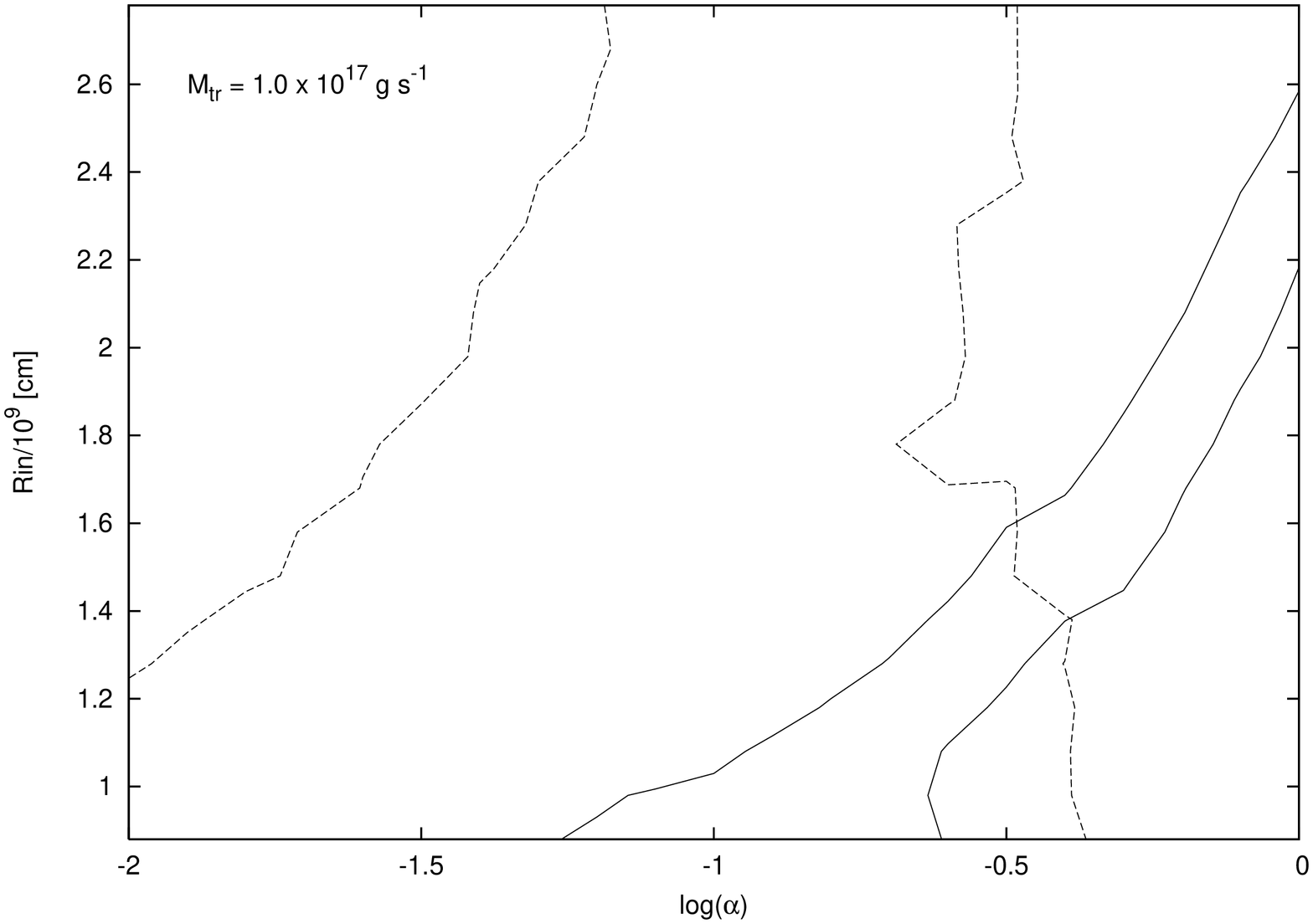}\\
\includegraphics[width=60mm,angle=-90]{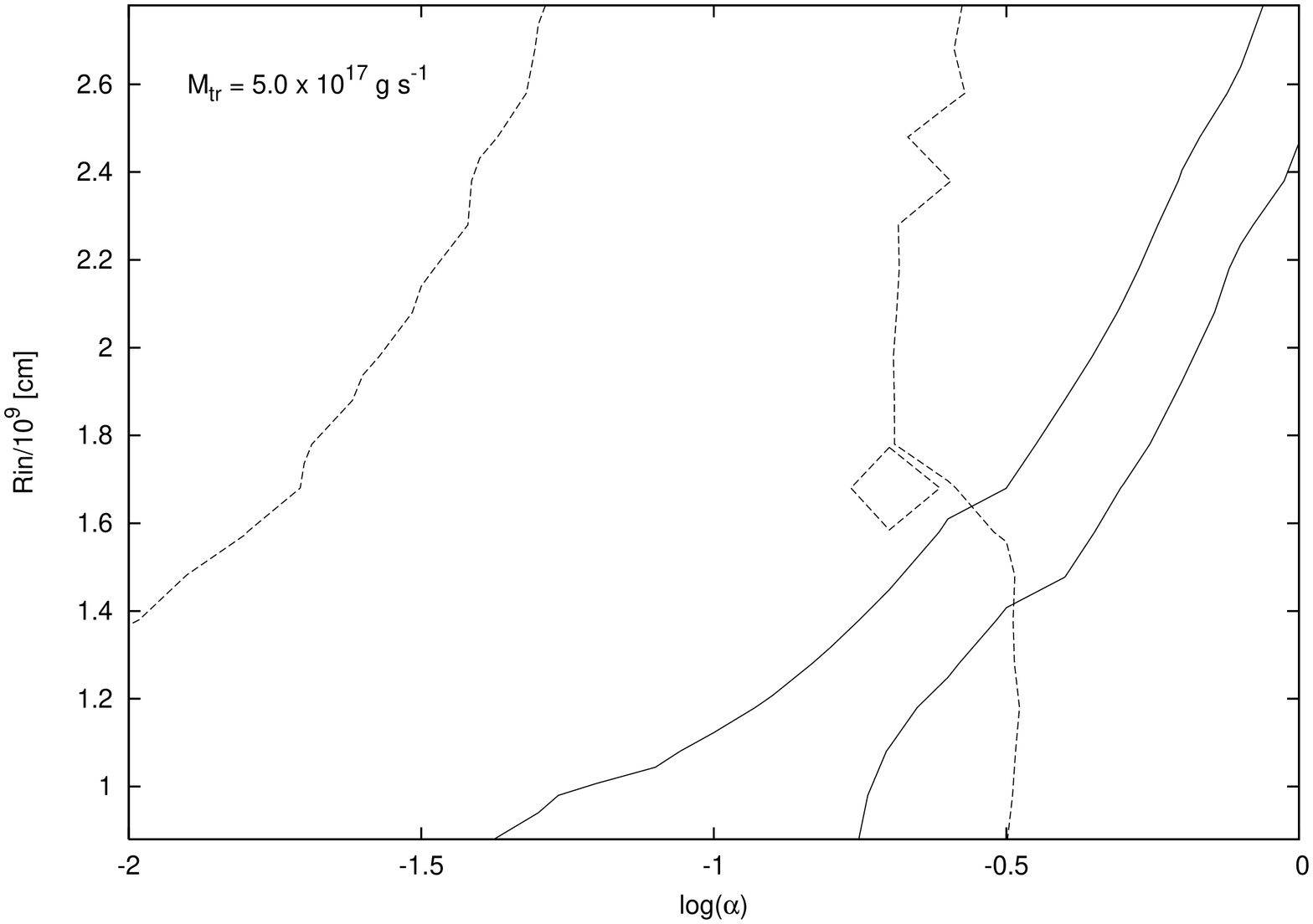}
\includegraphics[width=60mm,angle=-90]{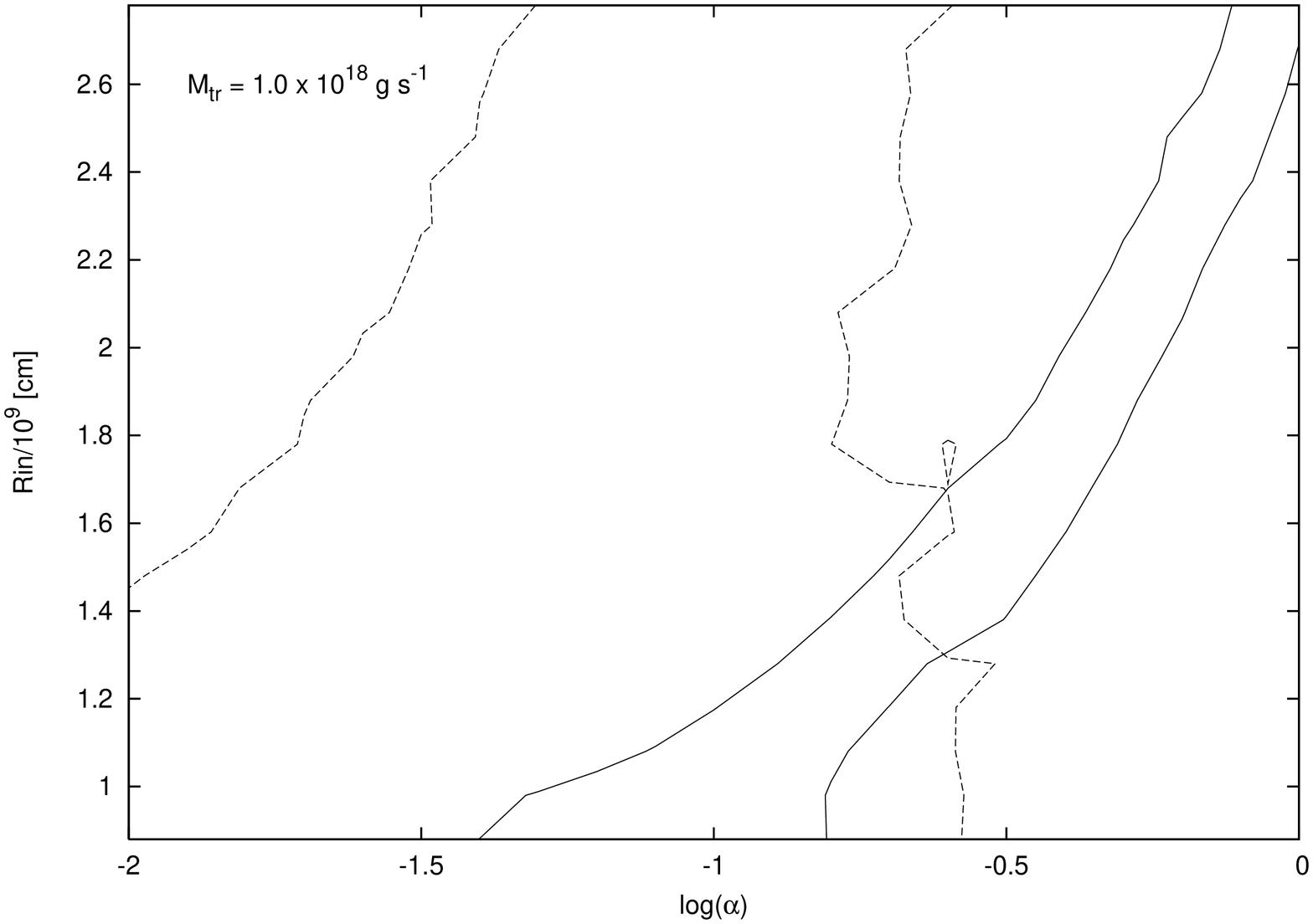}
\caption{Evolution of the $\pm 1$-$\sigma$ intervals of the power law (solid line) and the logarithm of the cut off frequency (dotted lines) for KR\,Aur in the parametric space $\alpha - R_{\rm in}$. Four different mass transfer rates are represented in each panel.}
\label{kraur_comp}
\end{figure*}

\begin{figure}
\includegraphics[width=60mm,angle=-90]{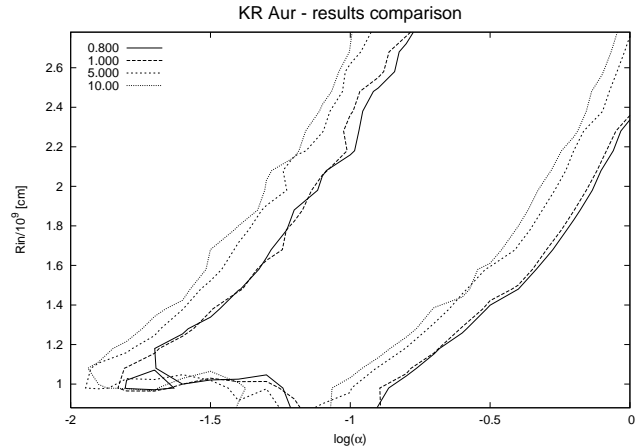}
\caption{Evolution of the power law index (right lines) and cut off frequency (left lines) for KR\,Aur as a function of mass transfer rate marked in units of $10^{17}$\,g\,s$^{-1}$.}
\label{results_comparison}
\end{figure}

The first clear result of our simulations is that the power law indexes measured from the PDS are in the $\alpha$ interval $0.1 - 1.0$ as required by hot discs of nova like systems in the high state. On the other hand, the cut off frequency points to low $\alpha$ values (see upper right panel in Fig.~\ref{results_grid}). Fig.~\ref{results_grid} also shows that the simulated power law index allows untruncated disc radii while the cut off frequency do suggests a truncated disc (i.e. $R_{\rm in}\ga 2.1 \times 10^{9}$ cm for $\alpha>0.1$). We then decided to search for $R_{\rm in}$ values in the range $\alpha = 0.1 - 0.5$ for which the power law index and cut off frequency overlap within the estimated errors and the results are listed in Table~\ref{results_tab_kraur}. Fig.~\ref{kraur_comp} shows the dependence of the PDS parameters with mass transfer rate in the $R_{\rm in} - \alpha$ plane. For the sake of clarity, Fig.~\ref{results_comparison} also shows the evolution of the mean PDS parameters without the error intervals. The displacement of the isolines towards lower $\alpha$ values with increasing mass transfer rate is clear.
%
%
%
%
\begin{table}
\caption{Extreme $\alpha$ and $R_{\rm in}$ values allowed by the overlap of the $\pm1 \sigma$ contours in Fig. 3. Different mass transfer rates are considered.}
\begin{center}
\begin{tabular}{lcccr}
\hline
\hline
$\dot{M}_{\rm tr}$ & min $\alpha$ & min $R_{\rm in}$ & max $\alpha$ & max 
$R_{\rm in}$ \\
($10^{17}$\,g\,s$^{-1}$) & & ($10^9$\,cm) & & ($10^9$\,cm) \\
\hline
0.800 & 0.10$^a$ & 0.88$^b$ & 0.40 & 1.56 \\
1.000 & 0.10$^a$ & 0.88$^b$ & 0.40 & 1.60 \\
5.000 & 0.10$^a$ & 0.88$^b$ & 0.33 & 1.64 \\
10.00 & 0.10$^a$ & 0.88$^b$ & 0.25 & 1.67 \\
\hline
\end{tabular}
\end{center}
$^a$ Minimum $\alpha$ parameter allowed by theory.\\
$^b$ Minimum inner disc radius as set by the radius of the white dwarf.\\
\label{results_tab_kraur}
\end{table}

\subsection{UU\,Aqr}
\label{results_uuaqr}

The results of our simulations for the case of UU\,Aqr are totally different than for the previous system. For illustration we show the case of the lower limit mass transfer rate ($8 \times 10^{16}$\,g\,s$^{-1}$) in the lower panels of Fig.~\ref{results_grid}. The observed cut off frequency $\log (f_{\rm ctf}) = 1.11$ is obtained for very low values of the $\alpha$ parameter beyond the upper left corner of the figure (where the simulated value is 1.2, i.e. minimal). Therefore, the whole interval lies outside the calculated $\alpha$ range. Assuming the same trend of $R_{\rm in}$ versus $\alpha$ as in the KR\,Aur case, very large inner disc radii would be required to get a physically realistic $\alpha$ parameter. Therefore, our simulations of the cut off frequency do not yield a satisfactory result. Similarly, the observed power law index can only be reproduced by our simulations for unplausible values $\alpha>1$. And we note that these conclusions do not depend on the assumed mass transfer rate.

The statistical distribution of the simulated cut off frequencies shown in Fig.1 (main figures) are different for KR\,Aur and UU\,Aqr. In the former case the histograms are well fitted with a Gauss function while in the latter case the Gaussian fit only gives a rough approximation. Consequently, our Gaussian-based uncertainties may not be very realistic. However, we note that a minimum error of 5-$\sigma$ is needed to bring the cut off frequency into required (plausible) $\alpha$ interval. On the other hand, the power law histogram clearly follows a Gaussian distribution and even a 10-$\sigma$ error is not able to bring the power law interval into acceptable $\alpha$ values in the UU\,Aqr case. Therefore, we conclude that our results are robust against a reasonable error amplification and hence no acceptable fit can be obtained for UU\,Aqr with our simulations.

\section{Discussion}
\label{discussion}

\subsection{Model degeneracies}

By fitting our turbulence model to the flickering data we find that the best fitted $\alpha$ parameter decreases while $R_{\rm in}$ increases with the selected value of $\dot{M}_{\rm tr}$. How can this be explained? The velocity of matter along the disc radius is given by (see e.g. \citealt{frank1992});
\begin{equation}
v_{\rm r} \sim \alpha^{4/5} \dot{M}_{\rm tr}^{3/10} M_{\rm 1}^{-1/4} r^{-1/4} f^{-14/5},
\label{disk_vr}
\end{equation}
where
\begin{equation}
f = \left[ 1 - \left( \frac{R_{\rm wd}}{r} \right)^{1/2} \right]^{1/4}.
\end{equation}
Equation \ref{disk_vr} shows that the radial viscous velocity $v_{\rm r}$ increases (and hence the flaring time scale decreases) with both $\dot{M}_{\rm tr}$ and $\alpha$. Therefore, the simulations performed with higher $\dot{M}_{\rm tr}$ result in a higher $v_{\rm r}$ and hence a shorter flaring time scale. As a result, the simulated statistics of the flickering changes and so do the PDS parameters, such as the power law slope and break frequency. In order to reproduce the observed power law properties, the drop in flickering time scale or event duration must be compensated by other parameters such as $\alpha$. In other words, the impact of increasing $\dot{M}_{\rm tr}$ can be compensated by decreasing the $\alpha$ parameter in order to keep the observed power law unchanged. This is the reason behind the $\dot{M}_{\rm tr}-\alpha$ correlation. A similar explanation holds for the $\dot{M}_{\rm tr}-R_{\rm in}$ behavior. If we increase $\dot{M}_{\rm tr}$ the characteristic time scale of the flickering events will drop and the modeled flickering statistics will contain a larger number of fast events. In order to eliminate the excess of fast events we need to cut down the inner disc, where short time scale events dominate. Therefore, we can compensate for the increase in $\dot{M}_{\rm tr}$ by increasing $R_{\rm in}$. A correlation between $\alpha$ and $R_{\rm in}$ was already found by \citet{dobrotka2010}. The explanation is again based on the dependence of the radial velocity on $\alpha$. By increasing $\alpha$ the radial velocity also increases which causes a decrease in the event duration. The quantity of short events is then enhanced and the inner disc radius must rise to compensate for it (see Dobrotka et al. 2010 for details).

\subsection{Results compared with DIM predictions}

The study of flickering using the statistical model of \citet{dobrotka2010} yield two different results for the nova like cataclysmic variables KR\,Aur and UU\,Aqr. A successful fit to the observed PDS properties is obtained for KR\,Aur but not UU\,Aqr. In the former case the observed power law index and cut off frequency can be reproduced with a truncated disc and parameter $\alpha \sim 0.1$, as required by the DIM (see \citealt{lasota2001} for review). Accretion disc theory sets a lower limit to $\alpha \simeq 0.1$ while the upper limit depends on the mass transfer rate used in the model. The higher the mass transfer rate, the lower the $\alpha$ parameter. Maximal values of the $\alpha$ parameter $0.25 - 0.40$ are obtained for mass transfer rates in the range $0.08 - 1 \times 10^{18}$\,g\,s$^{-1}$. These $\alpha$ values are in excellent agreement with the ones frequently used or derived in simulations ( $0.1 - 0.2$ , see e.g. \citealt{schreiber2003}, \citealt{schreiber2004}) and with the observed values $0.1 - 0.4$ of fully ionized geometrically thin discs (\citealt{king2007}). On the other hand, the inner disc radius is allowed to vary from nearly the white dwarf radius up to $1.56 - 1.67 \times 10^9$\,cm. The disc in KR\,Aur is found to be untruncated or only marginally truncated. The inner disc radius is found to increase with the mass transfer rate, but it is always considerably smaller than in the case of quiescent dwarf nova (see e.g. \citealt{schreiber2003}), when the disc is not in the steady state, i.e. the mass transfer rate at the inner disc radius is lower than the mass transfer rate from the secondary (i.e. at the outer disc). According to the DIM, the high mass accretion rate during dwarf nova outburst pushes the inner disc radius closer to the primary. The ionized steady state discs of nova like systems posses higher mass transfer rates than dwarf novae discs. Therefore, smaller inner disc radii are expected. The maximum inner disc radius allowed by our simulations ($1.56 - 1.67 \times 10^9$\,cm) are considerably smaller than the value 2.00 $\times 10^9$\,cm used for the dwarf nova SS\,Cyg in quiescence by \citet{schreiber2003}. This is in agreement with the standard accretion disc theory. In summary, our analysis has demonstrated that a simple statistical model of disc turbulence is able to reproduce the observed flickering characteristics of KR\,Aur with a set of disc parameters consistent with DIM theory. The turbulence scenario is then a plausible mechanism and no other processes seem to be required to explain the observed flickering activity.

The  case of UU\,Aqr is totally different to that of KR\,Aur. The observed cut off frequency lead to $\alpha$ values lower than 0.01 in the entire range of inner disc radii and for all mass transfer rates. This is unacceptable even for discs in the low state, such as dwarf novae in quiescence. Concerning the power law, our simulations yield to $\alpha$ values larger than the upper limit 1.0 which is not acceptable by basic accretion disc theory. Therefore, no suitable disc turbulence model can succesfully reproduce the observed PDS of UU\,Aqr.

To summarize, we have two similar nova like systems, but the analysis of their flickering properties with our turbulence disc model lead to two completely different results. Is the turbulence scenario wrong for the case of UU\,Aqr? \citet{baptista2008} found spiral structures in the eclipse maps of UU\,Aqr and they claim that the flickering is caused by turbulence generated by the interaction of the disc with the spiral density wave. The model of \citet{dobrotka2010} that we use in this work is based on a disc profile that does not take into account the presence of spiral structures. Furthermore, the model assumes that the largest dimension scale of the turbulent elements is set by the disc scale height. Any other structure like spiral arms can considerably affect the input distribution function of turbulence dimension. We propose that the presence of spiral structures generate turbulence with a totally different distribution function of dimension than in a standard disc and hence our model is not able to describe the situation. Therefore, the turbulence scenario cannot be tested in this system.

Are KR\,Aur and UU\,Aqr so different that one has spiral structures in the disc while the other has not? No eclipse maps of KR\,Aur are available for comparison but 
\citet{kato2002} did not find any evidence for superhumps. Superhumps are brightness variations modulated with a time scale close to the orbital period which are caused by disc precession. The physical phenomenon responsible for disc precession and superhump activity is the tidal force of the secondary star. Superhumps are triggered when the disc outer radius exceeds the 3:1 resonance radius. This is satisfied for mass ratios $q < 0.35$ (see e.g. \citealt{whitehurst1991}, \citealt{patterson2005}). Tidal forces do not only generate precessing discs and superhumps but also spiral shocks in the disc (see e.g. \citealt{steeghs2001}). According to the reported binary parameters UU\,Aur, with $q = 0.3$, is likely to be affected by the tidal influence of the secondary but not KR\,Aur with $q = 0.6$. Then different disc behavior may be expected in the two binaries.

\subsection{Distribution of flickering in the disc}

\begin{figure}
\includegraphics[width=59mm,angle=-90]{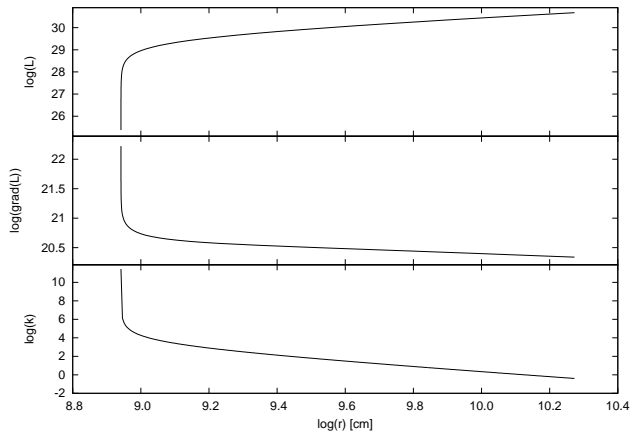}
\caption{Disc profile in KR\,Aur. Radial distribution of angular momentum (upper panel), gradient of angular momentum (middle panel) and correlation coefficient $k(r)$ (bottom panel), as computed from equation~\ref{correlation_coef}. The very steep variations at low $r$ values are due to the proximity of the white dwarf. Our simulations are computed at larger radii after the steep variations.}
\label{kraur_disc_prof}
\end{figure}

The parameter $k(r)$ provides an estimate of the number of events between two adjacent disc annuli and hence it is a prime indicator of flickering. The lower panel in Fig.~\ref{kraur_disc_prof} shows that $k(r)$ rises very steeply in the inner disc and decreases with increasing $r$. This suggests that the bulk of the events responsible for the flickering activity are located in the inner disc regions which is in agreement with the observations. \citet{dobrotka2010} explained this invoking the maximum dimension scale of the events. The inner disc contains very small turbulent eddies which are able to transfer little angular momentum. Therefore, a large number of eddies is needed. Conversely, the outer disc regions contain very large eddies which are able to transport large amounts of angular momentum. Therefore, a smaller number of events is sufficient.
This is best seen when inspecting the gradient of angular momentum rather than the angular momentum itself. The top panel in Fig.~\ref{kraur_disc_prof} clearly shows that the angular momentum increases toward the outer disc. Flickering is caused by the transport of angular momentum $L(r)$ and hence the difference between adjacent radii is what matters. The gradient of $L(r)$ decreases toward the outer disc, as is illustrated in the middle panel of Fig.~\ref{kraur_disc_prof}. Therefore, the amount of $L$ that needs to be transported decreases toward the outer disc and hence the quantity of turbulent eddies must also decrease (see lower panel in Fig.~\ref{kraur_disc_prof}). In summary, flickering becomes more important at low $r$ values or the innermost part of the disc, as observed (see Sec~\ref{introduction}).

\section{Summary}
\label{summary}

In this paper we have analyzed the flickering activity of two nova like systems KR\,Aur and UU\,Aqr. We applied a statistical model of flickering simulations developed by \citet{dobrotka2010} based on a simple idea of angular momentum transport in accretion discs. The use of a steady-state disc model requires the disc to be in the hot ionized state. The two selected nova like systems are then adequate targets for this analysis. We successfully fitted the observed power density spectrum of KR\,Aur using an inner disc truncation radius of $0.88 - 1.67 \times 10^9$\,cm and parameter $\alpha \sim 0.10 - 0.40$. Unfortunately, our modeling fails to reproduce the power density spectrum of UU\,Aqr, probably because of the presence of spiral structures detected by \citet{baptista2008}. Such structures are expected in the case of UU\,Aqr because of its low mass ratio $q = 0.3$. On the other hand, KR\,Aur has a larger mass ratio of 0.6 and the disc does not reach the 3:1 resonance radius where tidal forces from the secondary star generate spiral shocks. Therefore, a different disc behavior is naturally expected in the two systems.

Our simulations show that the number of flickering events increases at lower radii and hence concentrate in the inner disc regions. This behavior is explained by the gradient of angular momentum which also increases toward the center of the disc. Furthermore, the duration of the turbulence events depends on the radial viscous velocity which is a function of the mass transfer rate and the $\alpha$ parameter. All these parameters are correlated and, although our modelling provides useful constraints, they cannot be uniquely determined.

\section*{Acknowledgment}

This work is supported in part by the Grand-in-Aid for the global COE programs on "The Next Generation of Physics, spun from Diversity and Emergence" from MEXT. AD was supported also by the Slovak Grant Agency, grant VEGA-1/0520/10. Partly funded by the Spanish MEC under the Consolider-Ingenio 2010 Program grant CSD2006-00070: ''First science with the GTC'' (http://www.iac.es/consolider-ingenio-gtc/).

\vspace{-0.6cm}

\bibliographystyle{mn2e}
\bibliography{mybib}

\label{lastpage}

\end{document}